\documentclass[intlimits,twoside,a4paper]{article}
\usepackage[cp1251]{inputenc}

\usepackage[eqsecnum]{cmpj3}

\issue{2024}{27}{1}{13201}
\doinumber{10.5488/CMP.27.13201}

\title[Percolation connectivity in competitive RSA of binary disk mixtures]%
{Percolation connectivity in deposits obtained using competitive random sequential adsorption of binary disk mixtures%
\thanks{Dedicated to Prof. Jaroslav Ilnytskyi on the occasion of his 60th birthday}}
\author[N. I. Lebovka, M. R. Petryk, N. V. Vygornitskii] {
N. I. Lebovka\orcid{0000-0002-8314-0601}\refaddr{label1}  \thanks{Corresponding author: \email{lebovka@gmail.com}.},
M. R. Petryk\orcid{0000-0001-6612-7213}\refaddr{label2},
N. V. Vygornitskii\orcid{0000-0003-0982-8878}\refaddr{label1}
}
\addresses{
\addr{label1} Laboratory of Physical Chemistry of Disperse Minerals, F. D. Ovcharenko Institute of Biocolloidal Chemistry of NAS of Ukraine, Kyiv 03142, Ukraine
\addr{label2} Ternopil Ivan Puluj National Technical University, 56 Ruska Street, Ternopil 46001, Ukraine 
}

\date{Received July 08, 2023, in final form September 6, 2023}

\begin{document}

\maketitle

\begin{abstract}
Connectedness percolation phenomena in the two-dimensional (2D) packing of binary mixtures of disks with different diameters were studied numerically. The packings were produced using random sequential adsorption (RSA) model with simultaneous deposition of disks. The ratio of the particle diameters was varied within the range $D=1$--$10$, and the selection probability of the small disks was varied within the range 0--1. A core-shell structure of the particles was assumed for the analysis of connectivity. The packing coverages in a jamming state for different components, connectivities through small, large and both types of disks, the behavior of electrical conductivity were analyzed. The observed complex effects were explained accounting for the formation of conductive ``bridges'' from small disks in pores between large disks. 

\printkeywords
\end{abstract}

\section{Introduction\label{sec:intro}}

Nanoporous and nanostructured materials and films have many useful modern applications in different adsorption, catalytic, electro-magnetic, microelectronics, colloid lithography, storage, optical, and biomedical devices \cite{Plawsky2009,Leclerc2012, Petryk2015,Soonmin2019,Zighem2021}. The model of random sequential adsorption (RSA) is widely used to mimic properties of such materials. The basic variant of RSA assumes a random and irreversible adding  of non-overlapping particles to a system (they do not overlap with any other previously deposited particles), and very strong binding of the particles at the deposition place without any detachment and diffusion effects. Various variants of RSA models were extensively discussed in different reviews \cite{Evans1993, Meakin1993, Talbot2000, Adamczyk2017, Adamczyk2022}.

For RSA deposition on two-dimensional (2D) substrate, the so-called ``jamming limit'' corresponds to the saturation coverage of the surface $\varphi$. For a simple one-component model without any interactions between particles and post deposition diffusion processes, the value $\varphi$ mainly depends on the shape of deposited particles. Particularly, for identical disks $\varphi=0.5471$ \cite{Finegold1979, Tanemura1979, Feder1980, Hinrichsen1986, Hinrichsen1990, Wang1994,Talbot2000, Huang2023}.
This value of $\varphi$ is much smaller than those observed for a random loose packing ($\varphi_{RLP} \approx 0.66$--0.67), random close packing ($\varphi_{RCP} \approx 0.82$--0.83), and maximum value of coverage for the disks arranged in a triangular lattice ($\varphi_{max} = \piup/2\sqrt{3} \approx 0.907$) \cite{Brouwers2023}. 

The geometrical properties of the 2D RSA configurations of identical disks were analyzed in detail~\cite{Hinrichsen1986}. The `direct' and `indirect' neighbors, and `stable' and `unstable' holes associated with them were identified. The parabolic shape of the hole size distribution was revealed with the hole diameter $h$ restricted within $0.1537< h/d < 1$, where $d$ is a disk diameter, and $1/\sin(\piup/3) - 1\approx 0.1537$ is an Apollonian ratio. Later on, the 2D RSA configurations of equal-sized disks were compacted up to $\varphi \approx 0.772$ and their geometrical properties were also analyzed \cite{Hinrichsen1990}. These random packings were in relatively stable configurations. 2D RSA problems were also investigated for other shapes, e.g., for deposition of ellipses~\cite{Talbot1989, Sherwood1990, Ricci1992, Viot1992, Talbot2000, Haiduk2018, Abritta2022}, rectangles \cite{Vigil1989, Vigil1990, Ricci1992, Viot1992, Talbot2000}, unoriented squares \cite{Viot1990}, discorectangles~\cite{Vigil1989, Ricci1992, Viot1992, Talbot2000, Haiduk2018, Lebovka2020}, needles and fibers \cite{Sherwood1990, Tarjus1991a, Ricci1992, DeBianchi2021, Chao2022, Uetsuji2022}, squares \cite{Feder1980, Viot1990}, regular polygons \cite{Ammi1987}, and many other complex shapes \cite{Zhang2013, Haiduk2018, Ciesla2020, Ciesla2022, Kubala2022, Morga2022}.

For deposition of disks with different sizes, the most popular are simultaneous and multi-step RSA models. In simultaneous models, the disks are selected for trial deposition onto the substrates with some predefined probability, and in the multi-step models the substrates is covered by the disks with different sizes sequentially, e.g., preliminary by smaller disks and then larger ones up to the complete jamming state.

The simultaneous RSA deposition models for disks with different sizes were discussed in many works~\cite{Talbot1989, Talbot1994,Meakin1992,Wagaskar2020}.
The theory of RSA deposition  for a binary mixture of hard disks of greatly differing diameters was developed \cite{Talbot1989}. The jamming limits for the larger disks $\varphi_D$ at different diameter  ratios $D/d$~($=5,10, \infty$) were estimated. The 2D RSA deposition of polydisperse disk mixtures with a continuous distribution of sizes was also theoretically studied \cite{Tarjus1991}.  
The simultaneous RSA deposition model for disks with two sizes or distribution of sizes onto a 2D planar substrate were investigated using a computer simulation approach \cite{Meakin1992}. 
For the binary mixture, the size ratio $D/d$ was within the range $1.125<D/d<8$ and the 
fraction of small disks $p$ selected for a trial deposition (selection probability) was within the range $0.016<p<0.875$. Quite different deposition kinetics for large and small disks was observed. In saturation state, the total surface coverage $\varphi_T$ and coverage for large disks $\varphi_D$ decreased with increasing the fraction of small disks $p$, whereas the coverage for small disks $\varphi_d$ increased with increasing $p$. In the limit $p\to 0$ and $D/d\to\infty$,  the total and small-disks coverages were estimated to be $\varphi_T=0.7945$ and $\varphi_d=0.2478$, respectively. These simulation data were in satisfactory agreement with theoretical results obtained in the limit of $D/d\to\infty$ \cite{Talbot1989}. 
The data on surface coverages $\varphi_d(D)$ and $\varphi_D(D)$ at different selection probabilities $p$ revealed rather intriguing facts. Particularly, at $p=0.5$ and $p=0.875$,  significant minima in the behavior of $\varphi_D(p)$ were observed at $D\approx 1.5$. The reasons for such a behavior were not discussed. 

The simultaneous RSA deposition model for binary disk mixtures has been also recently discussed in detail \cite{Wagaskar2020}. The studies were performed for size ratios within $D/d=1.11$--10 and for different relative deposition rate constants. In the systems studied,  the total surface coverage $\varphi_T$ was always greater than 0.547, and for a given value $D/d$, the maximum of $\varphi_T$ was observed at some optimum value of the  relative rate constant. 

The simultaneous RSA deposition model for  polydisperse mixture of disks on decorated substrates was also studied \cite{Marques2012}. The polydispersity was simulated using a truncated Gaussian-size distribution, and adsorption was allowed within the (equal) square cell patterns. Different distinct regions (interacting and non-interacting cell-cell adsorption, and single-particle-per-cell and multi-particle-per-cell adsorption) were identified in dependence on cell-cell separation and cell size. The 2D RSA deposition of binary mixtures of variously shaped oriented Lam\'{e} 
super-disks was recently discussed \cite{Svrakic2016}.
The best values of saturation coverages were observed for the chosen objects with the similar shapes.

The 2D RSA deposition processes on heterogeneous, partially pre-covered or pre-patterned surfaces (multi-step RSA models) were also studied in many works 
\cite{Adamczyk1997,Adamczyk1998,Adamczyk1998a,Adamczyk2001,Adamczyk2001a,Adamczyk2002,Adamczyk2002a,Weronski2005,Weronski2007,Weronski2007a,Araujo2008,Sadowska2021}.
The deposition of disks on different 2D fractal surfaces (Vicsek fractal, Serpinski triangle and squares fractals)  was also discussed~\cite{Ciesla2012,Tartaglione2021}. 

In multi-step RSA models, the initial adsorption of small disks at first step  significantly affected the deposition of large disks and diminished their adsorption rates at the second step \cite{Adamczyk1997,Adamczyk1998} (for a more detailed analysis and review of these works, see 
\cite{Adamczyk2005,Adamczyk2012,Adamczyk2017,Adamczyk2022}). 

Note, that rather interesting data were obtained for 2D random packing of binary hard discs using gravity protocol \cite{Barker1989,Odagaki2002,Okubo2004}. The discs were selected with probability $p$, dropped vertically using the rolling down procedure, and adsorption was completed when the disc reaches a local potential energy minimum. The structural phase diagram was represented as the total packing coverage $\varphi_T$ versus the $D/d$ dependence \cite{Barker1989}. The functions $\varphi_T(D/d)$ passed the minimum at some value of $D/d$ that was dependent on the value of $p$.  The dependencies of the total packing coverage $\varphi_T$ versus the area fraction of smaller discs $n=\varphi_d/\varphi_T$ at $n\approx 0.35$ showed maximum $\varphi_T \approx 0.854$ for $D=10$ and minimum $\varphi_T \approx 0.804$ for $D=1.11$ \cite{Odagaki2002,Okubo2004}. Self-organization in binary mixtures of disks was recently studied using 2D Monte Carlo simulation of dead-end diafiltration process \cite{Lebovka2022}. Stratification of the disks in vertical direction was observed and incomplete separation of mixture was explained by the formation of an impenetrable barrier from larger particles at the deposit bottom.

The connectivity and percolation properties of heterogeneous systems composed of different species are very important for understanding of electrical conductivity, flow and transport properties of such materials \cite{Sahimi2023}. In previous studies, the connectivity problems were mainly discussed for 2D deposits of fully penetrable discs \cite{Quintanilla2007,Balram2010,Mertens2012,Speidel2018}. Particularly, the surface coverage at the percolation threshold for identical discs was estimated to be $\varphi_p=0.676(3)$ \cite{Quintanilla2000,Zuyev2003}. For binary dispersions of disks the higher percolation thresholds were observed than for monodisperse disks \cite{Meester1994,Meester1996}. 

Moreover for systems with distribution of disk sizes, the surface coverage at percolation threshold showed a non-trivial dependence on the distribution \cite{Phani1984}. For a given value of the disk diameter ratio $(D/d=1/7-12.5)$, the maximum of surface coverage at the percolation threshold $\varphi_p$ was observed at the intermediate fraction of small disks $p$ \cite{Quintanilla2001,Quintanilla2006,Quintanilla2007}. Particularly, at $D/d=10$ and $p=0.99$, the estimated value $\varphi_p \approx 0.75981$ was much higher compared with the surface coverage for a monodisperse system ($\varphi_p \approx 0.676339$ at $p=1$). 

The morphology and continuous percolation in porous films simulated by using the RSA deposition of disks on 2D substrate was also recently discussed \cite{Janssens2022}. The 2D percolation and cluster structure of the compact random packing of binary disks  were studied \cite{He2002}. The packings with different diameter ratio $D/d=1$--5 and relative compositions were generated using the repeated relaxation and expansion procedures to obtain the overlap-free packings. This allowed to generate of relatively compact structures with direct contacts between particles. The disk size ratio $D/d$ significantly affected the percolation value for the area fraction $\varphi_d$ of small disks, and the increase in $D/d$ leads to a decrease in $\varphi_d$. 

In the pure RSA packing, there are practically no direct contacts, and the hopping transport is only possible for the particles covered by tunneling shells. This corresponds to the so-called connectedness percolation of non-overlapping particles with a core(hard)--shell(soft) structure \cite{Drwenski2018,Lebovka2020}. At the percolation point, the complete electrical path through the systems is formed.  For RSA deposits of identical disks with no direct contacts,  it was mentioned that the increase of the disk radius by 20\% results in the formation of the percolating cluster of overlapping particles \cite{Hinrichsen1986}. More recently, the connectivity analysis for RSA deposits formed by identical discorectangles with a core-shell structure was performed~\cite{Lebovka2020,Lebovka2021, Lebovka2021a}. 
For identical disks in a jamming state, the estimated relative reduced thickness of the shell was $\delta/d \approx 0.0843 \pm 0.001$. However, as far as we know, the percolation connectivity of 2D RSA configurations for a competitive model in the mixture of impenetrable disks had never been discussed before.

\begin{figure}[htb]
	\centerline{\includegraphics[width=0.45\textwidth]{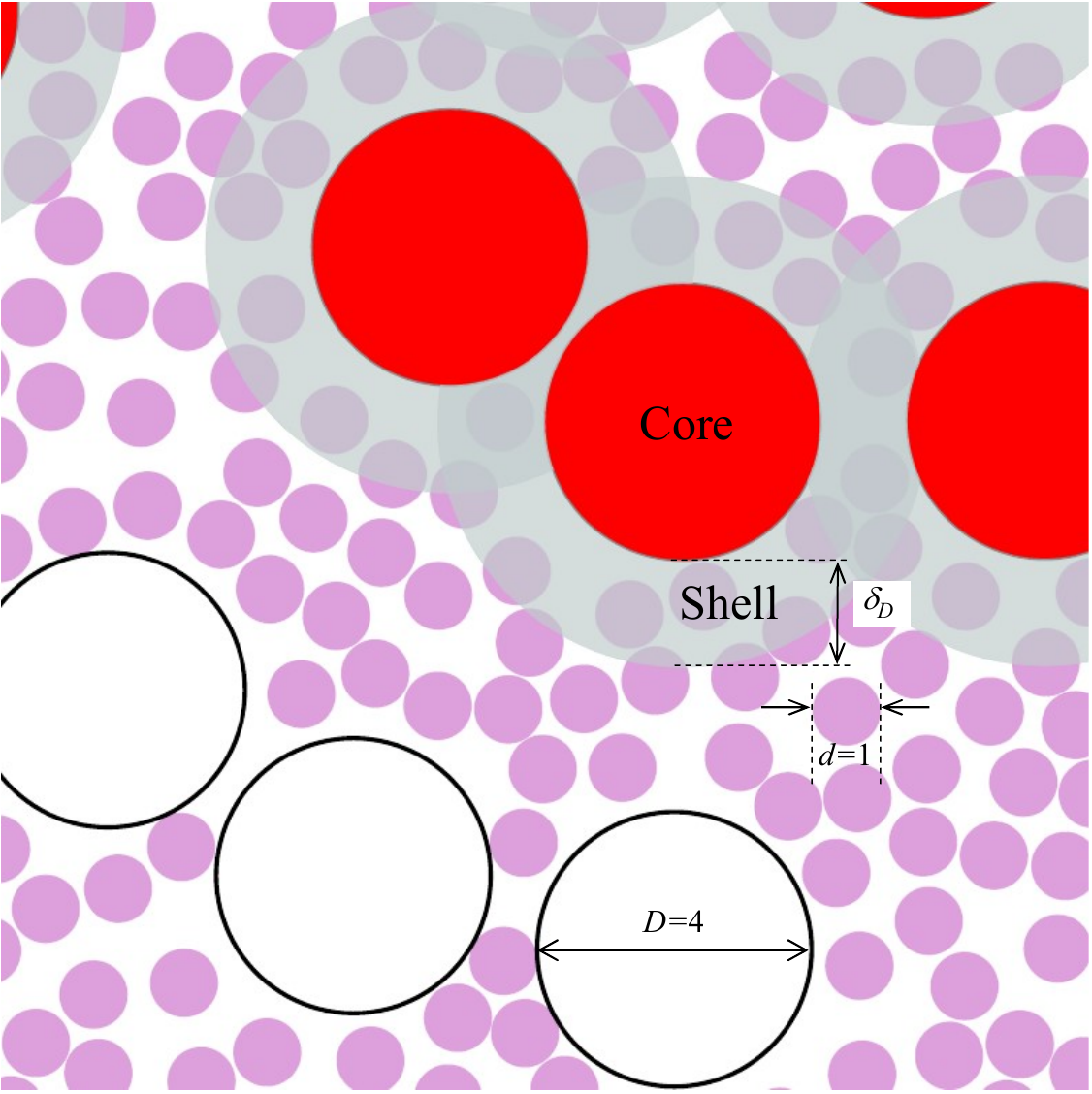}}
	\caption{(Colour online) An illustration of the RSA packing of binary mixture in jamming state on a 2D substrate. The total size of the system was $128\times 128$ and only a part with the size of $16 \times 16$ is shown. Intersections of the disk cores are forbidden. The soft (penetrable) shell with thickness $\delta_D$ is shown only for large disks (colored) belonging to the percolation cluster. The data are presented for the diameter of large disk $D=4$ and selection probability $p=0.5$.} \label{fig_Model}
\end{figure}

This work discusses the percolation connectivity in 2D RSA deposits obtained for binary disk mixtures with the application of a simultaneous RSA deposition model. The structure of this work is as follows. Section II presents the computational technical details and the main definitions. Section III presents the main results, and the final section IV summarizes our conclusions.

\section{Computational model\label{sec:methods}}
The 2D deposits were formed using the algorithm of the RSA model similar to that described previously \cite{Meakin1992,Wagaskar2020}. The binary mixtures of disks with diameters $d$ and $D$ ($d \leqslant D$) were deposited randomly and sequentially. At each deposition attempt, the small disc was selected with the probability $p$ (hereinafter referred to as the selection probability) and the large disc was selected with the probability $1-p$. The overlapping with previously deposited disks was forbidden. In this work, the lengths of all objects are presented in units of the smaller disk diameter  $d$.  To reduce the finite effects, the periodic boundary conditions along the $x$ and $y$ directions were used. 

\begin{figure}[htb]
	\centerline{\includegraphics[width=0.45\textwidth]{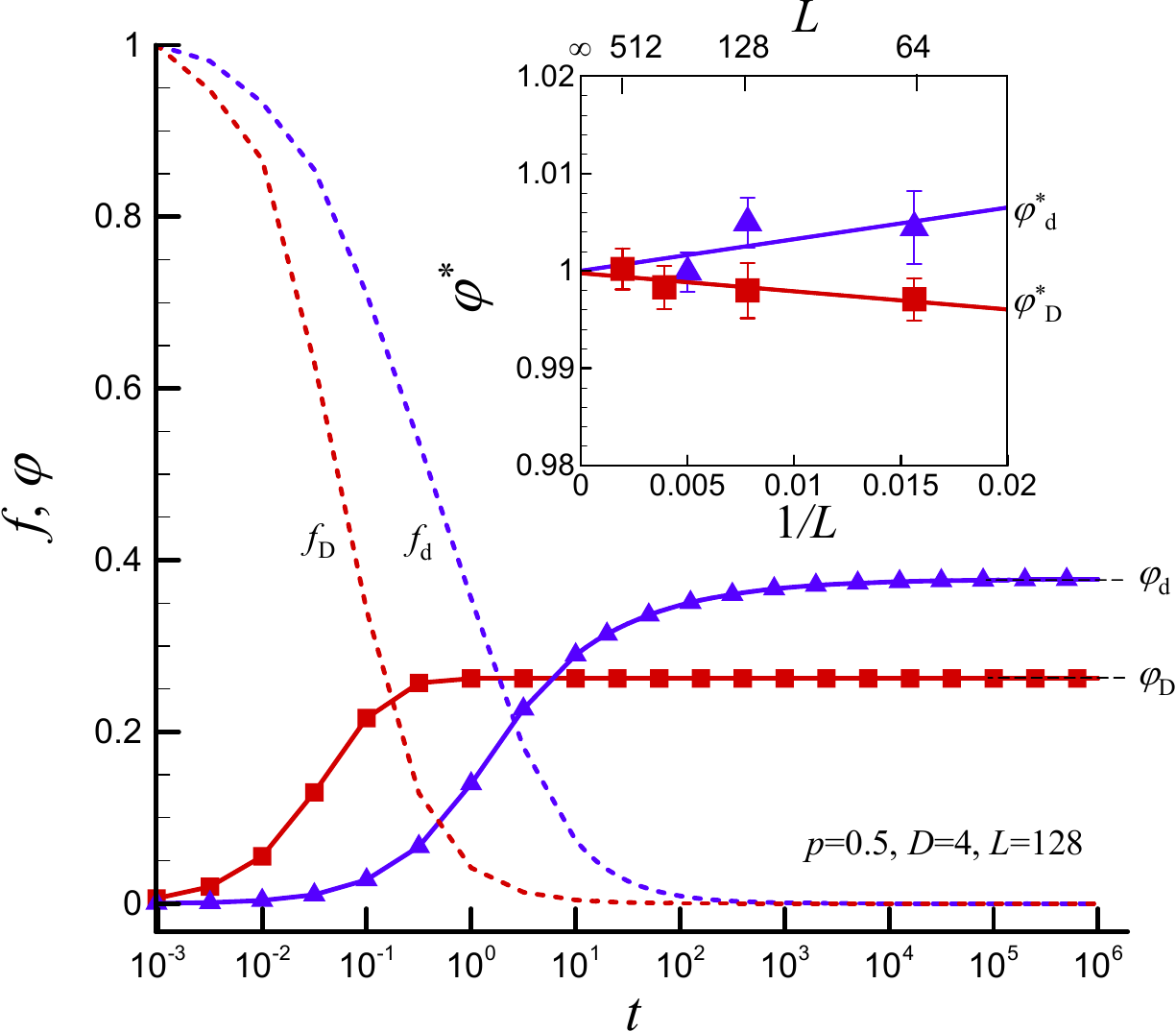}}
	\caption{(Colour online) Example of time dependencies of the surface coverages $\varphi$ and deposition probabilities $f$ for small and large discs. The data presented for a particular case with the diameter of large disk $D=4$, selection probability $p=0.5$, and system size $L=128$. Here, $\varphi_d$, $\varphi_D$ are the jamming coverages for small and large disks in the limit $t \to \infty$. Inset shows finite-size scaling dependencies of the normalized values of jamming concentrations $\varphi^*_d=\varphi_d/\varphi_d(L\to\infty)$, $\varphi^*_D=\varphi_D/\varphi_D(L \to \infty)$ within the range $L=64$--512.} \label{fig_Kinetic_Scaling}
\end{figure}
Figure \ref{fig_Model} presents an example of the 2D RSA packing of binary mixture in a jamming state obtained for the diameter of large disk $D=4$ and the selection probability $p=0.5$. The small part of the system with the size of $16 \times 16$ is presented.
The percolation connectivity of the system in the jamming state was analyzed using the core–shell model as described earlier \cite{Lebovka2020}. The connectivity of discs was tested checking the overlapping of circular outer soft shells around the hard disks.

The minimum (critical) value of shell thickness required for the formation of spanning clusters along the $x$ or $y$ direction, was evaluated using the lists of near-neighbor disks and the Hoshen--Kopelman algorithm \cite{Hoshen1976,Marck1997}.
Analysis was performed for determination of the thickness of shells for individual connectivity through small ($\delta_d$) or large ($\delta_D$) disks and simultaneous connectivity through small and large disks ($\delta_T$). In the analysis of individual connectivity, the shells cover small ($\delta_d$) or large ($\delta_D$) disks. For example, the connectivity through large disks is shown in figure~\ref{fig_Model}. The conductive shells with thickness $\delta_D$ cover only large disks (colored) belonging to the percolation cluster. In the analysis of simultaneous connectivity through small and large disks, the shells of the same thickness $\delta_T$ cover both small and large disks.

The electrical conductivity $\sigma $ of deposits was evaluated using a supporting square mesh of size $m \times m$ ($m=2048$). The mesh cells located at the core, shell, or pore parts were associated with extremely high ($\sigma_c=10^{12}$), intermediate ($\sigma_s=10^6$) and low ($\sigma_p=1$) electrical conductivities. In the absence of direct contacts between cores and for large contrasts between electrical conductivities, the main contribution to the electrical conductivity is expected to be from the overlapping of shells. For the calculation of electrical conductivity, the Frank--Lobb algorithm based on the $Y-\Delta$ transformation was used \cite{Frank1988}. More details can be found elsewhere \cite{Lebovka2021}. 

The time duration of the deposition process was calculated as $t =N/L^2$, where $N$ is the number of deposition attempts, and $L$ is the linear size of the system in $x$ and $y$ directions \cite{Lebovka2020a, lebovka2023two}. The surface coverages were evaluated as 
\begin{subequations} \label{eq:varphi} 	\nonumber
	\begin{align}
     \varphi_d & =n_d \piup d^2/4L^2, \\
	 \varphi_D & =n_D \piup D^2/4L^2,
	\end{align}
\end{subequations}
where $n_d$ and $n_D$ are the numbers of the deposited small and large disks, respectively. 

The deposition probabilities for small and large discs ($f_d$ and $f_D$, respectively) were evaluated accounting for the fraction of the successful disk deposition attempts. For particular species, these values were calculated as the ratios of successful disc deposition attempts to the total disc deposition attempts.

Figure \ref{fig_Kinetic_Scaling} shows typical time dependencies of the surface coverages $\varphi$ and deposition probabilities $f$ for small and large discs. The data are presented for a particular case with $D=4$ and $p=0.5$. The surface coverages $\varphi$ for both disks increases with time and approaches their jamming values $\varphi_d$, $\varphi_D$ in the limit $t \to \infty$. The deposition probabilities $f_d$, $f_D$ decrease with $t$ and gradually approach zero in the limit $t \to \infty$. In many cases, the deposition of large discs was rapidly blocked by the smaller discs. Typically at a relatively large time, the RSA deposition of small disks in the holes between large disks was observed. 

In order to evaluate the finite size scaling effects, preliminary tests were performed for the systems with different values of $L=128$--1024. The inset to figure~\ref{fig_Kinetic_Scaling} shows an example of the scaling behavior of jamming concentrations normalized values $\varphi^*_d=\varphi_d/\varphi_d(L \to \infty)$, $\varphi^*_D=\varphi_D/\varphi_D(L \to \infty)$ within the range $L=64$--512. Particularly, for $L=128$, the deviations of $\varphi^*_d$ and $\varphi^*_D$ from unity were smaller than $1\%$.

\begin{figure}[htb]
	\centerline{\includegraphics[width=0.45\textwidth]{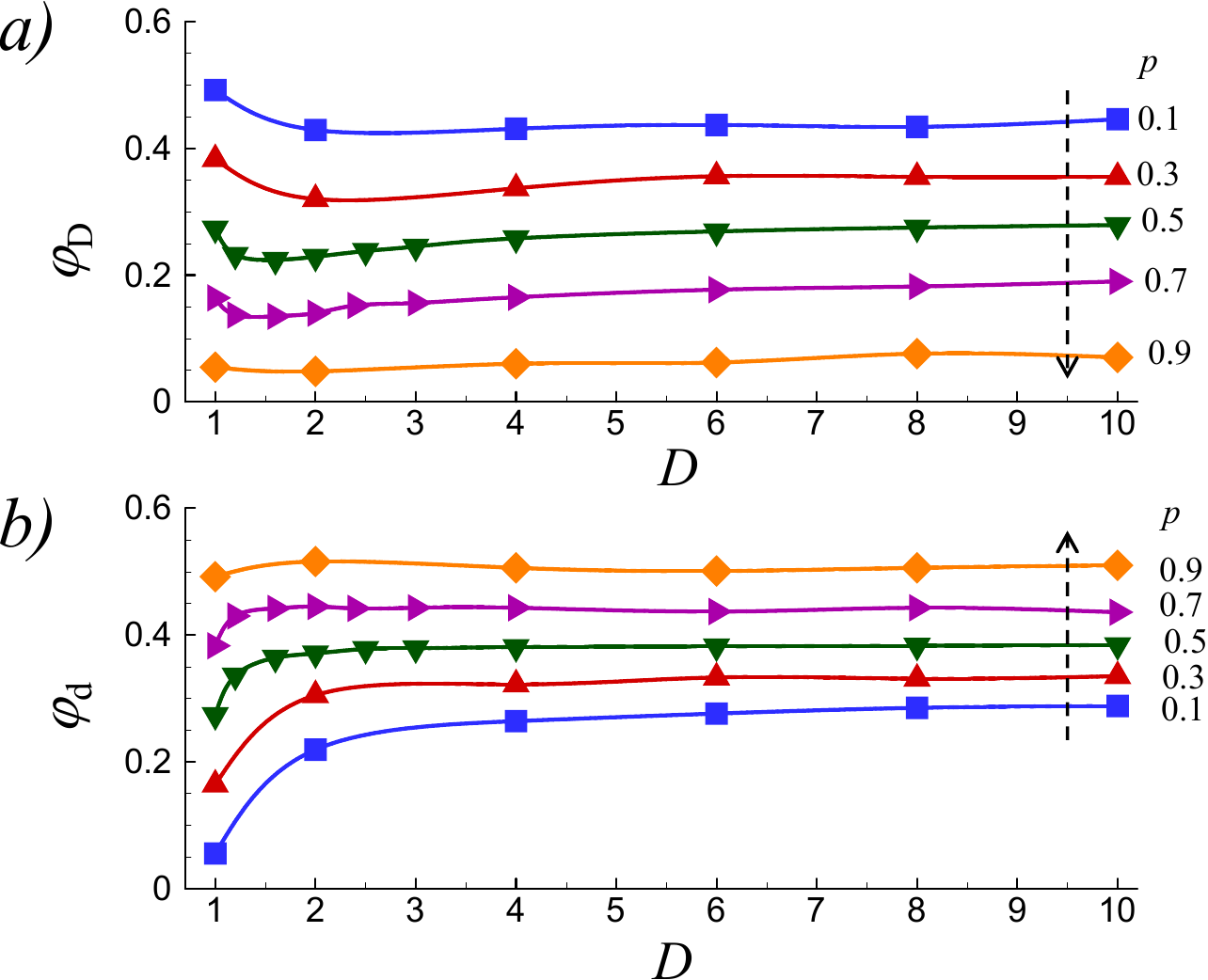}}
	\caption{(Colour online) Jamming coverages $\varphi_D$ (large disks) (a) and $\varphi_d$ (small disks) (b) versus diameter of large disk $D$ at different values of selection probability $p$. } \label{fig_Jamm_vs_D_for_p}
\end{figure}

The computer experiments were repeated using from $10$ to $1000$ independent runs. The error bars in the figures correspond to the standard deviations of the means at a significance level of 0.05. When not shown explicitly, they are of the order of the marker size.

\section{Results and discussion\label{sec:results}}

Figure \ref{fig_Jamm_vs_D_for_p} presents the behavior of jamming coverages for large $\varphi_D$ (a) and small $\varphi_d$ (b) disks  at different values of selection probability $p$. For large disks, the minima in $\varphi_D (D)$ dependencies were observed at some intermediate value of $D$ ($D \approx 1.5$--2.5) whereas for small disks, the $\varphi_d$ values continuously grew with $D$ increasing. With $p$ increasing, the opposite behavior in jamming concentrations for large and small disks was observed, in particular, $\varphi_D$ decreased and $\varphi_d$ increased. This behavior clearly reflects the increase in blocking of the surface by small disks as $p$ increases. 
  
\begin{figure}[htb]
	\centerline{\includegraphics[width=0.45\textwidth]{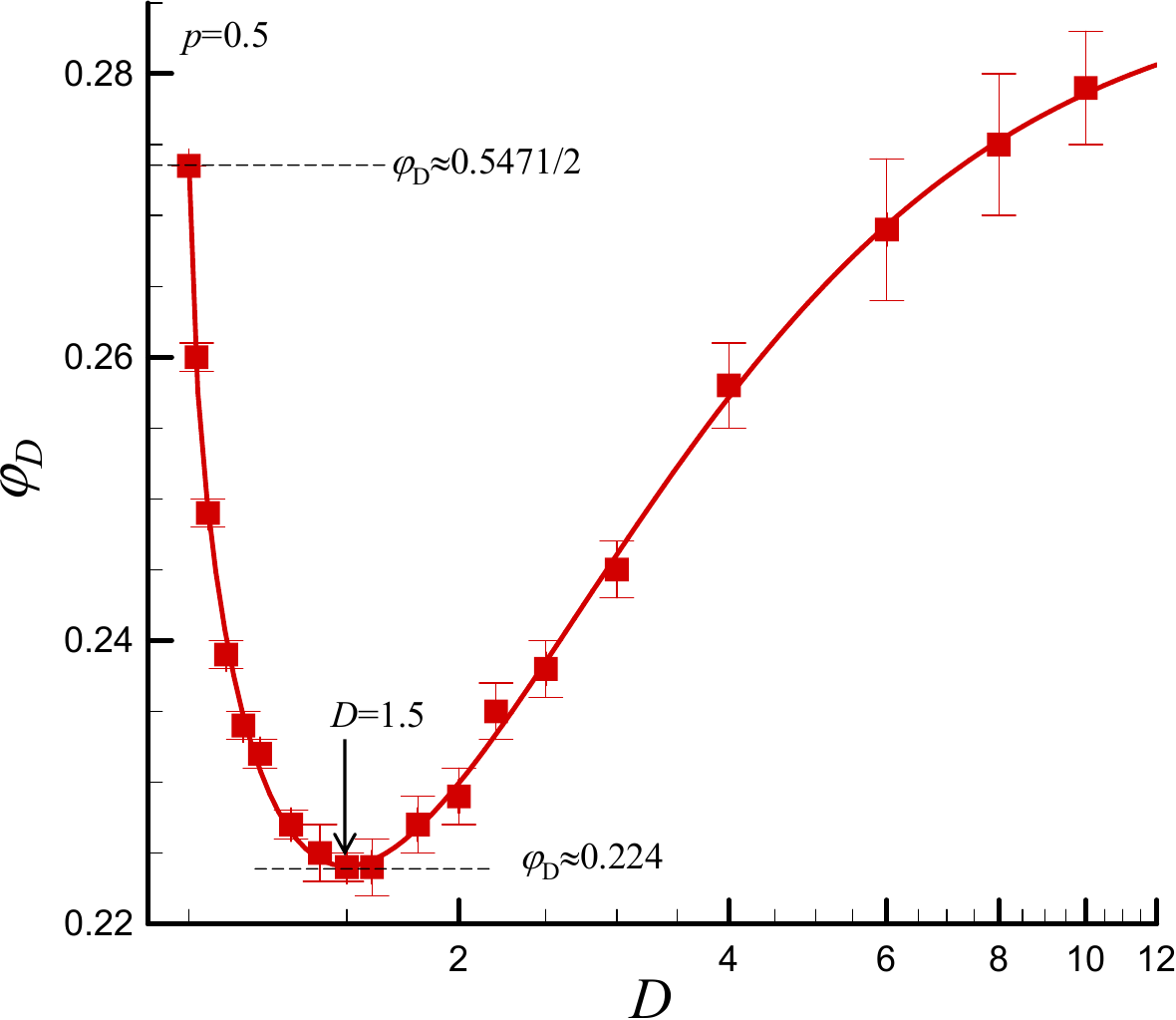}}
	\caption{(Colour online) Example of large particle jamming coverage $\varphi_D$ versus diameter of large disk $D$ for  the selection probability  $p=0.5$. } \label{fig_phiD_vs_D_for_p05}
\end{figure}
\begin{figure}[h]
	\centerline{\includegraphics[width=0.45\textwidth]{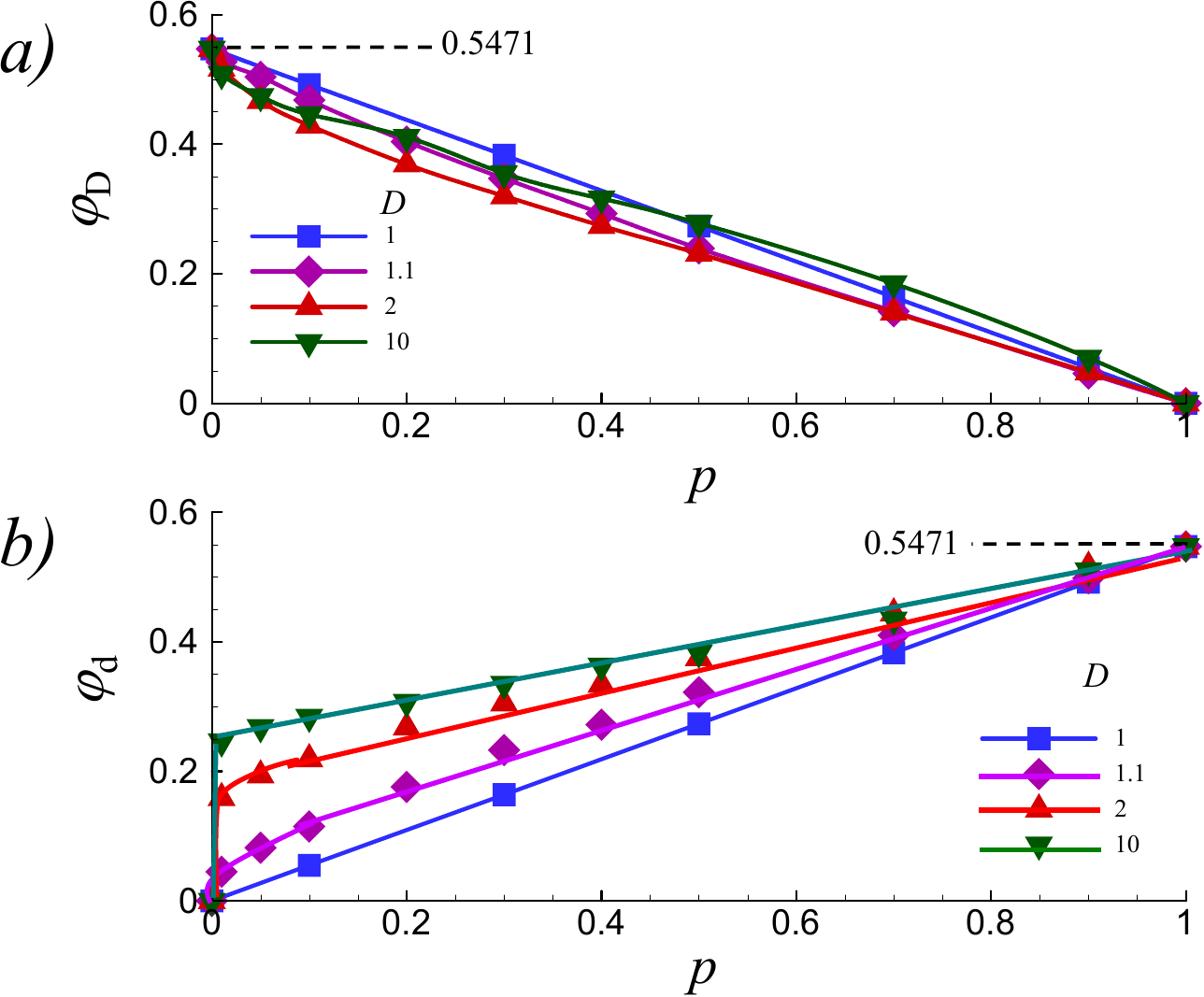}}
	\caption{(Colour online) Jamming coverages $\varphi_D$ (large disks) (a) and $\varphi_d$ (small disks) (b) versus the selection probability $p$ at different diameter of large disks $D$. The dashed lines at $0.5471$ corresponds to the jamming coverage of the identical disks \cite{Evans1993}.} \label{fig_Jamm_vs_p}
\end{figure}

Figure \ref{fig_phiD_vs_D_for_p05} presents more clear visualization of such a minimum extreme in $\varphi_D(D)$ behavior for a particular case $p=0.5$.  For $D=1$, we have  $\varphi_D \approx 0.274$, that is, about half of the value jamming coverage for identical disks $\varphi \approx 0.547$ \cite{Evans1993}. The minimum of $\varphi_D \approx 0.224$ was observed at $D \approx 1.5$. 

\begin{figure}[htb]
	\centerline{\includegraphics[width=0.45\textwidth]{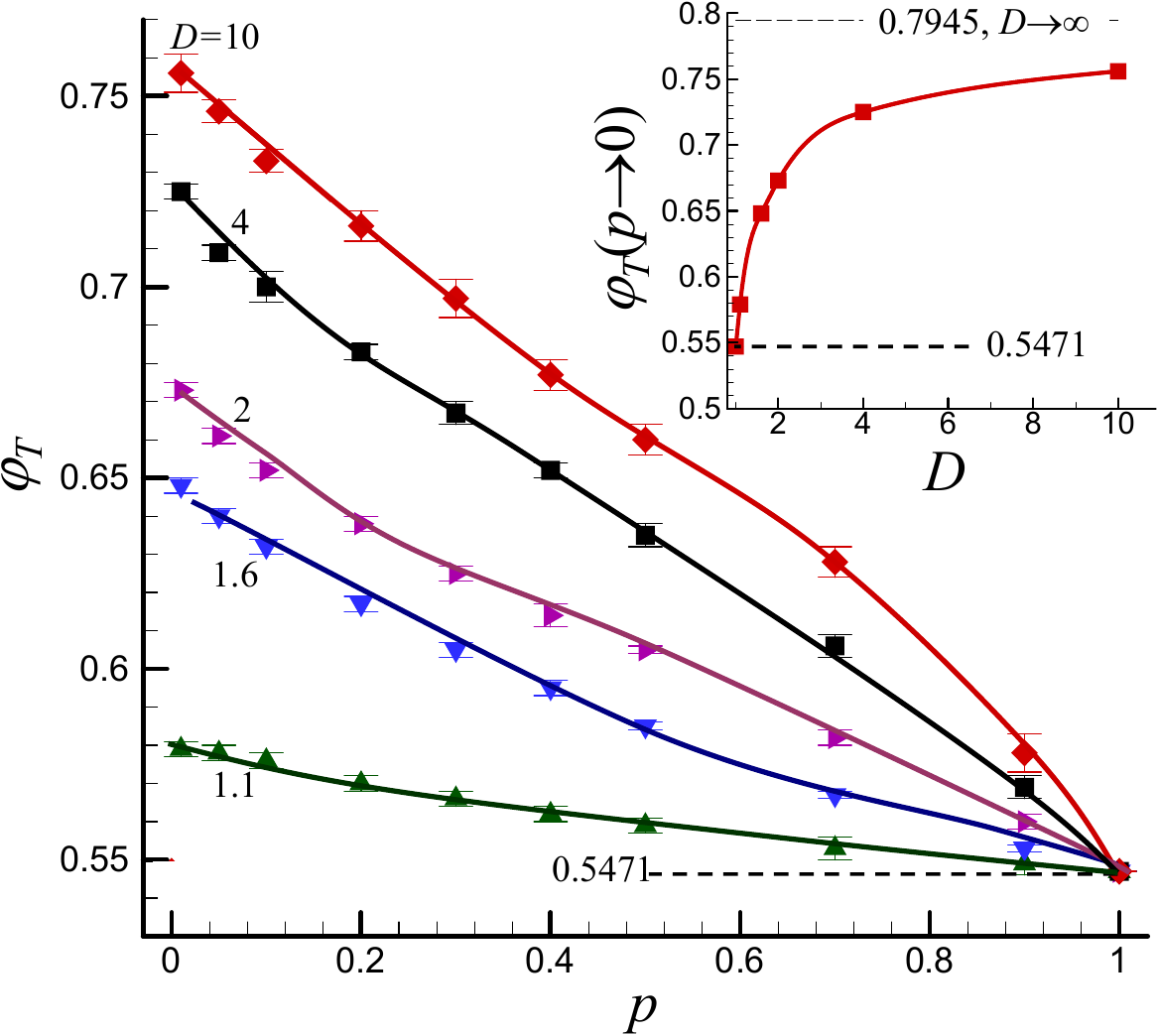}}
	\caption{(Colour online) The total jamming coverage $\varphi_T$ ($=\varphi_D+\varphi_d$) versus the selection probability $p$ at different diameter of large disks $D$. The dashed lines at $0.5471$ corresponds to the jamming coverage of the identical disks \cite{Evans1993} and at $\varphi_T=0.7945$ corresponds to the maximum possible value in the limit $D\to\infty$, $p \to 0$ \cite{Meakin1992}. } \label{fig_JammTotal_vs_p}
\end{figure}

Similar observations were reported earlier \cite{Meakin1992}, although explanations of such striking effects are still absent. This evidently reflected the relationships between the surface coverages and deposition probabilities~$f$ of different disks. At $p=0.5$, a strong competition in the deposition between disks of different size exists. For disks with approximately the same size ($D<1.5$), the initial decrease in $\varphi_D$ was accompanied by a more intensive increase in $\varphi_d$. It is interesting to note that the minimum extreme behavior was also observed for a quite different problem related with the regular compact packing of binary mixture of disks for a small difference between the sizes of disks (at $D \approx 1.5685d$) \cite{Kennedy2006}. 

For large $D$, our simulation data correlate well with the theoretical estimates obtained for $D \gg 1$~\cite{Talbot1989}. Particularly, at $p=0.5$ and $D=10$, we have $\varphi_D=0.280 \pm 0.004$, whereas the theory predicted $\varphi_D\approx 0.29$ for the same  values of $p$ and $D$.

Figure \ref{fig_Jamm_vs_p} presents the jamming coverages $\varphi_D$ (large disks) (a) and $\varphi_d$ (small disks) (b) versus the  selection probability $p$ at a different diameter of large disks $D$. The limiting values at $p=0$ (large disks, figure~\ref{fig_Jamm_vs_p}a) and at $p=1$ (small disks, figure~\ref{fig_Jamm_vs_p}b) correspond to the adsorption of identical disk with the jamming coverages $0.547$. For large disks, the values of $\varphi_D$ continuously decreased with increasing of $p$, and approximately linear $\varphi_D(p)$ dependencies were observed (figure~\ref{fig_Jamm_vs_p}a). At $D=1$, i.e., for disks with identical diameters, the predictable linear increase of $\varphi_d$ with increasing of $p$ was observed. However, at $D>1$, the initial jumps in $\varphi_d(p)$ at a  very small $p$ were always observed. These jumps show that the formation of jamming configurations for large disks with $\varphi_D \approx 0.5471$ was followed by the packing of small disks in the holes between large disks. For example, at $D>10$, the jump up to $\varphi_d  \approx 0.246$ was followed by practically linear increase of $\varphi_d \approx 0.246$  with $p$ ($\varphi_d  \approx 0.246 + 0.301 p$). The observed behavior was in good correspondence with the previously reported data \cite{Meakin1992}.

\begin{figure}[htb]
	\centerline{\includegraphics[width=0.9\textwidth]{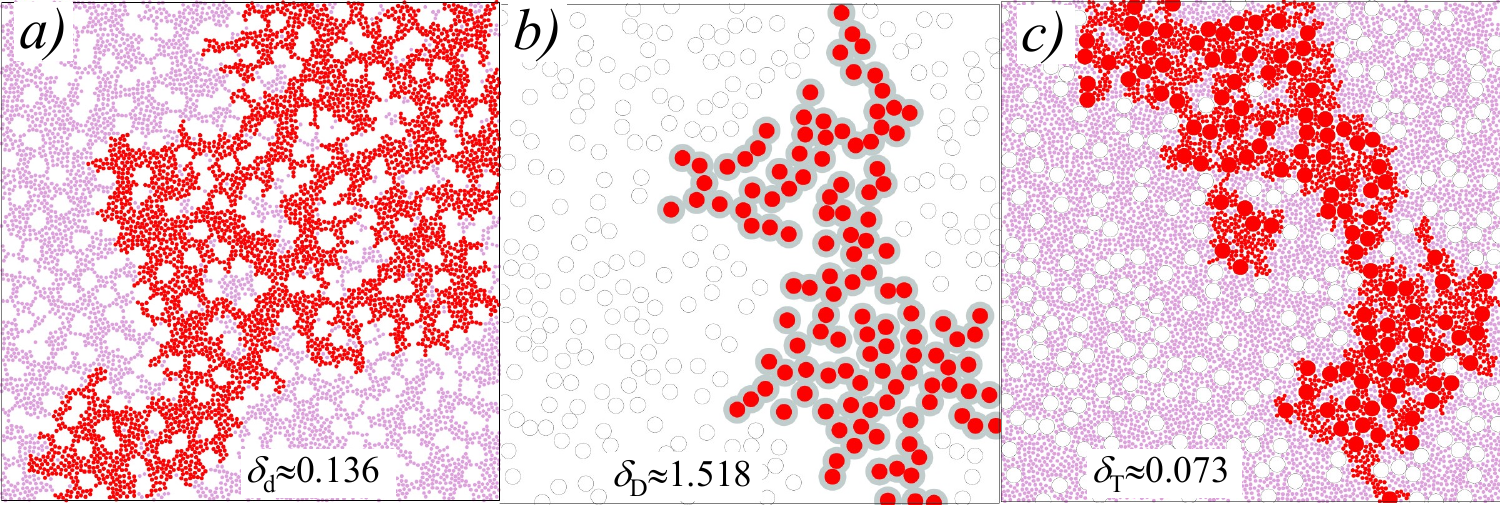}}
	\caption{(Colour online) Examples of percolation clusters (colored) for connectivities through the small~(a), large (b) and all (c)	(small+large) disks in the system. The patterns are presented for the system size $128\times 128$, selection probability  $p=0.5$, and diameter of large disks $D=4$. The presented values ($\delta_d$), ($\delta_D$), and ($\delta_T$) represent the mean values of the shell thicknesses  around the small, large, and both small and large disks, correspondingly.} \label{fig_PClusters}
\end{figure}
Figure~\ref{fig_JammTotal_vs_p} presents the total jamming coverage $\varphi_T$ ($=\varphi_D+\varphi_d$) versus the selection probability $p$ at different diameter of large disks $D$. In all cases, a continuous decrease of $\varphi_T$ with $p$ was observed.  Moreover, at fixed $p$, the values of $\varphi_T$ increased with an increase of $D$. The inset to figure~\ref{fig_JammTotal_vs_p} presents an example of such a behavior $\varphi_T (D)$ in the limiting case $ p \to 0$. Here, the value $\varphi_T$ varies between $0.5471$ (the jamming coverage of the identical disks \cite{Evans1993}) and $\varphi_T=0.7945$ (the maximum possible jamming coverage in the limit $D \to \infty$)~\cite{Meakin1992}. Therefore, for simultaneous 2D RSA deposition of binary mixtures, even at very small values of $p$, the total coverage contains a significant input from small disks. Particularly, in the limit $D \to \infty$, the input of small disks is relatively large $\varphi_d \approx 0.2474$. This input reflects the RSA packing of small disks in the pores between extremely large disks.  


Figure \ref{fig_PClusters} presents examples of connectivity analysis for models with shells around small (a, $\delta_d$), around large (b, $\delta_D$) and around both small and large (c, $\delta_T$) disks. In these models, the electrically conductive small (figure~\ref{fig_PClusters}a), large (figure~\ref{fig_PClusters}b), and both small and large (figure~\ref{fig_PClusters}c) disks were covered by electrically conductive shells. 

\begin{figure}[htb]
	\centerline{\includegraphics[width=0.5\textwidth]{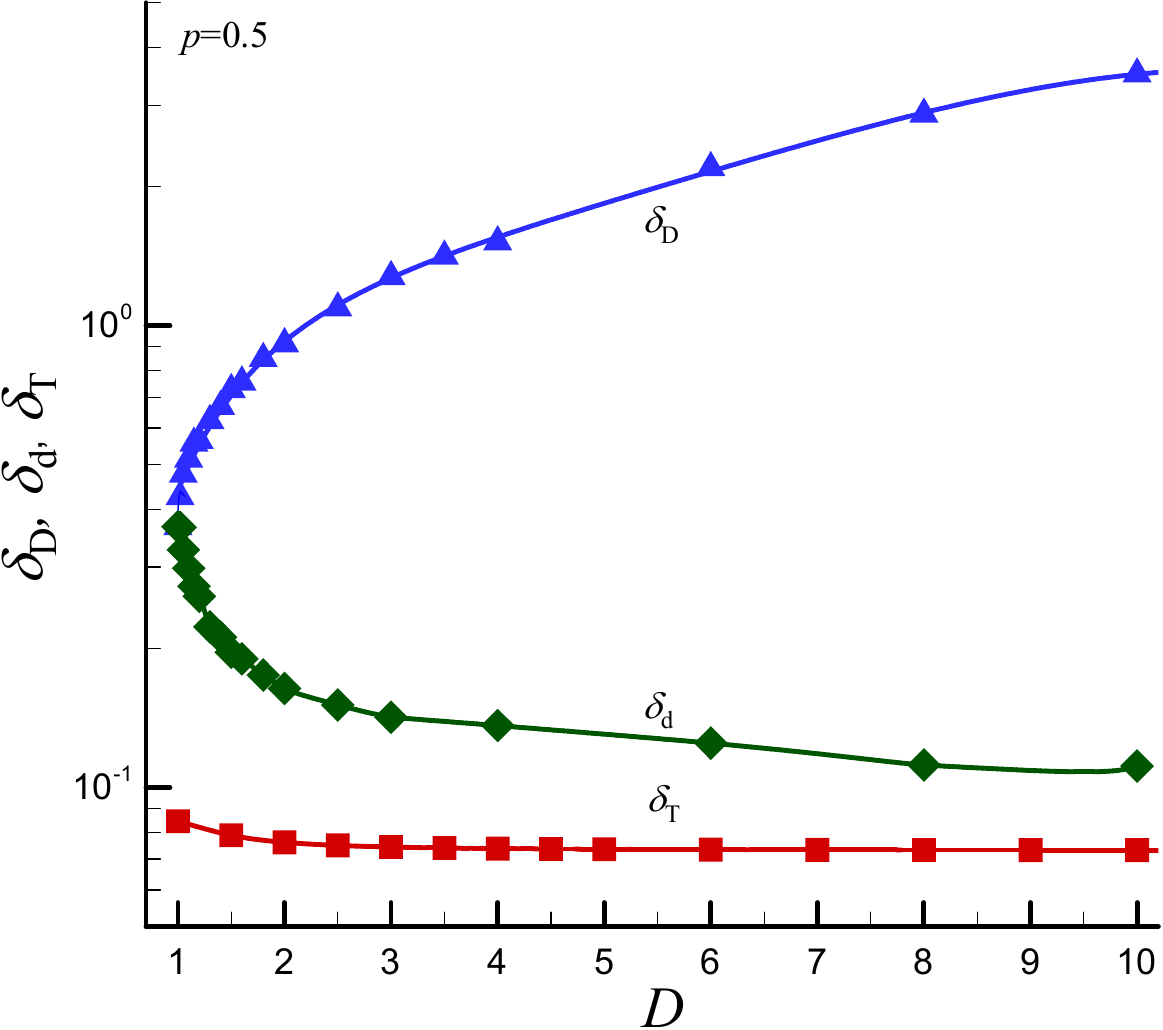}}
	\caption{(Colour online) Shell thickness around small $\delta_d$, around large $\delta_D$ and around both small and large $\delta_T$ disks versus the diameter of large disks $D$ at a fixed selection probability $p=0.5$.} \label{fig_Delta_vs_D}
\end{figure}

\begin{figure}[htb]
	\centerline{\includegraphics[width=0.5\textwidth]{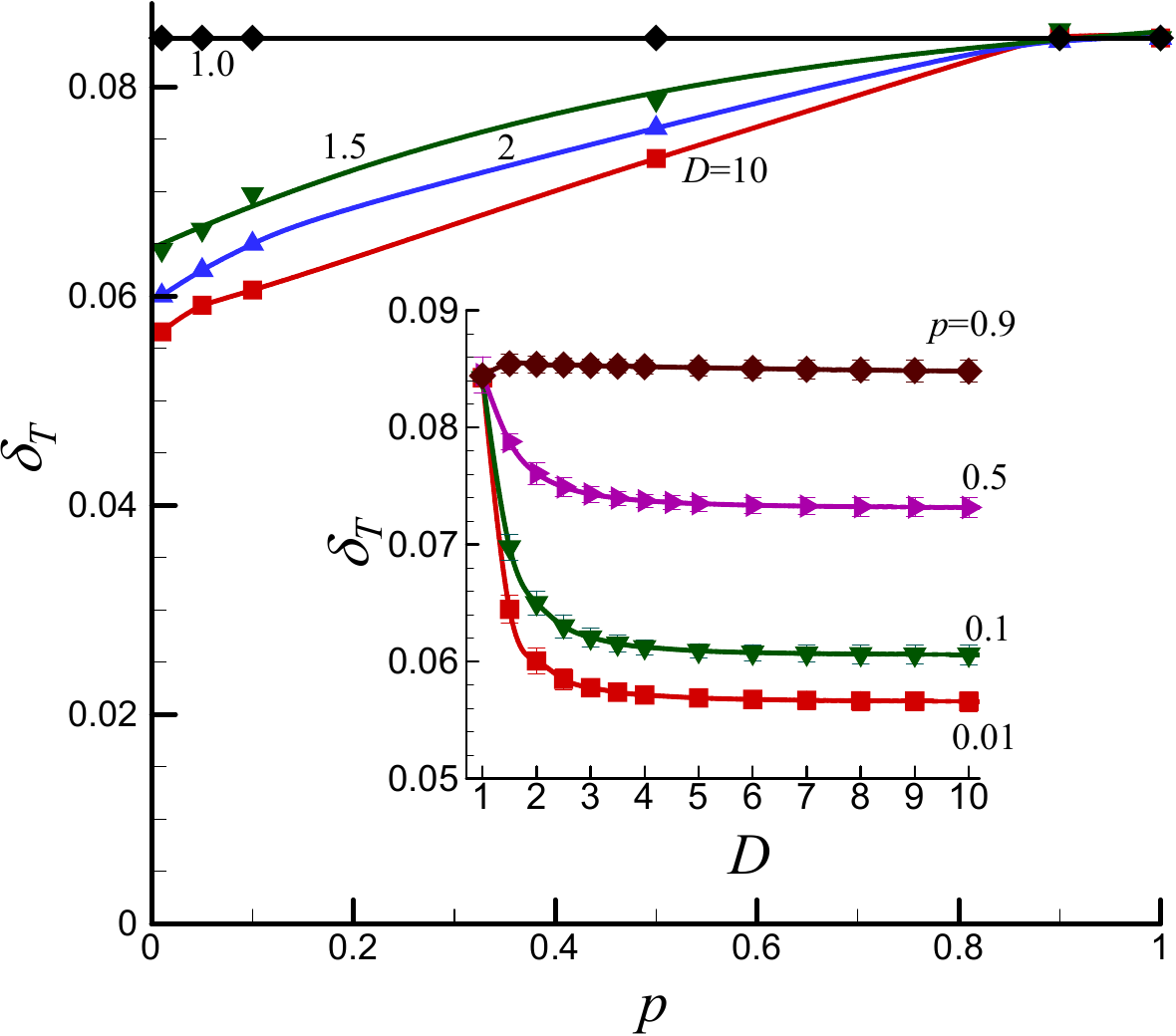}}
	\caption{(Colour online) Shell thickness around both small and large $\delta_T$ disks versus the selection probability $p$ at different values of large disk diameter $D$. The inset shows examples of $\delta_T(D)$ dependencies at different values of $p$.} \label{fig_DeltaT_behav}
\end{figure}
Examples of   $\delta_d (D)$, $\delta_D (D)$ and $\delta_T (D)$ dependencies at a fixed selection probability $p=0.5$ are presented in figure~\ref{fig_Delta_vs_D}. For one component 2D RSA packing of disks, the thickness of the percolation shell was recently estimated to be $\delta=0.084 \pm 0.001$ \cite{Lebovka2020}. For binary mixtures, the shell thickness for connectivity through both disks $\delta_T$ slowly decreased with increasing of $D$ and reached $\delta_D\approx 0.077$ at $D=10$. The shell thicknesses for the individual connectivity through small $\delta_d$ and large $\delta_D$ coincide at $D=1$ ($\delta_d=\delta_D=0.367\pm 0.001$). With increasing of $D$ the opposite behaviors of $\delta_d$ and $\delta_D$ were observed. The values of $\delta_d$ decreased and the value of $\delta_D$ increased with increasing of $D$ and reached the levels of $\delta_d \approx 0.111$ and $\delta_D \approx 3.504$ at $D=10$.

Figure \ref{fig_DeltaT_behav} presents the shell thickness around the both small and large $\delta_T$ disks versus the selection probability $p$ at different values of a large disk diameter $D$. In these mixtures, the value of $\delta_T$ was noticeably smaller than the shell thickness for a one component system $\delta=0.084 \pm 0.001$ \cite{Lebovka2020} and values of $\delta_d$ and $\delta_D$. Moreover, the values $\delta_T$ increased with increasing of $p$. The inset shows examples of $\delta_T(D)$ dependencies at different $p$. The values $\delta_T$ decreased with increasing of $D$ and significant changes in $\delta_T$ were only observed at relatively small values of $D$ ($D<3$). 

\begin{figure}[htb]
	\centerline{\includegraphics[width=0.5\textwidth]{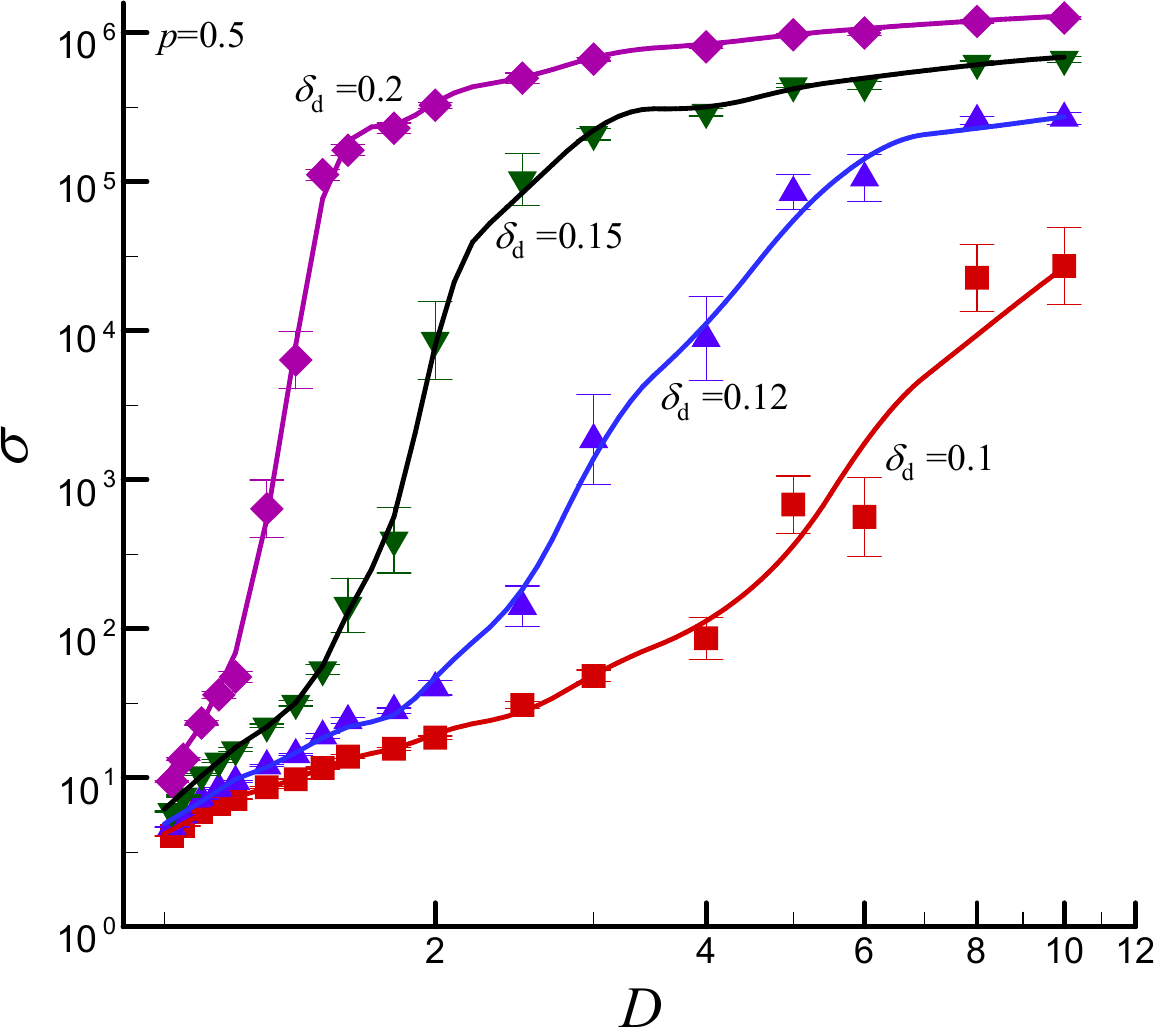}}
	\caption{(Colour online) Electrical conductivity $\sigma $ versus the large disk diameter $D$ for different values of the shell thickness for small disks $\delta_d$. The example of dependencies for a fixed value of the selection probability $p=0.5$ is presented. It was assumed that the non-conductive large disks without shells have small electrical conductivity (the same as for pores, $\sigma=1$), and the small disks are electrically conductive ($\sigma_c=10^{12}$ for cores and $\sigma_s=10^6$ for shells).} \label{fig_Conductivity}
\end{figure}
Note that at $D>2$, the values of $\delta_T(D)$ were relatively small at rather small values of $p$ (e.g., $p=0.01$ in the inset to figure~\ref{fig_DeltaT_behav}). This can be explained by the effective deposition of small disks in pores between large disks and by the formation of some kind of ``bridges'' effectively connecting the large disks. Note, that such electrical conductivity behavior was observed in segregated multi-component composite materials \cite{Lebovka2006}. In such systems, the small particles form conductive ``bridges'' on the surface of large particles and this resulted in the improved electrical conductivity. 

In order to demonstrate the impact of such effects, we evaluated the electrical conductivity of 2D RSA deposits with conductive small disks with shells and isolating large disks without shells. Figure \ref{fig_Conductivity} shows the electrical conductivity $\sigma$ versus the large disk diameter $D$. Note that at $D>2$ the surface coverage by smaller conductive disks was approximately the same, $\varphi_d \approx 0.38$ (figure~\ref{fig_Jamm_vs_D_for_p}b). However, the electrical conductivity of the whole system was greatly affected by the both $\delta_d$ and  $D$ values. 

\section{Conclusions}
The percolation connectivity in jammed systems obtained using competitive 2D RSA deposition for a two-component mixture of hard disks has been studied. The main parameters of the model are the ratio of particle diameters $D$ and selection probability $p$. The simulations were performed at $D=1$--10 and $p=0$--1. The detailed investigations of packing coverages in a jamming state for small $\varphi_d$ and large $\varphi_D$ particles were performed. The behaviors of the shell thicknesses required for the individual connectivity through the small $\delta_d$, large $\delta_D$ disks and through all the disks in the system $\delta_T$ were analyzed. 
Particularly, at $D>2$, the values of $\delta_T(D)$ were relatively small at rather small values of $p$ (e.g., $p=0.01$ in the inset to figure~\ref{fig_DeltaT_behav}). These effects were explained by the effective deposition of small disks in the pores between large disks and by the formation of some kind of electrical ``bridges'' effectively connecting the large disks. The natural extension of this work would be to explore the binary mixtures with varying shapes and/or sizes of the depositing particles. Further analysis of such a  behavior will be the subject of the forthcoming papers.

\section*{Acknowledgements}
We acknowledge the funding from NASU (KPKVK No 7.4/3-2023, 6541230, N.L.), Ministry of Education and Science of Ukraine (No~DI~247-22, 0122U001859, M.P.), and National Research Foundation of Ukraine (No. 2020.02/0138, N.V.V.).

\bibliographystyle{cmpj}
\bibliography{MS}

\ukrainianpart

\title{Перколяційна зв'язність в конкурентній моделі випадкової послідовної адсорбції бінарних упаковок дисків}
\author{М. І. Лебовка \refaddr{label1}, М. Р. Петрик\refaddr{label2}, М. В. Вигорницький\refaddr{label1}}
\addresses{
\addr{label1} Лабораторія фізичної хімії дисперсних мінералів Інституту біоколоїдної хімії імені Ф. Д. Овчаренка НАН України, Київ 03142, Україна
\addr{label2} Тернопільский національний технічний університет імені Івана Пулюя, вул. Руська, 56, Тернопіль 46001, Україна 
}

\makeukrtitle

\begin{abstract}
\tolerance=3000%
Методом комп’ютерного моделювання досліджено перколяційну зв’язність двовимірних (2D) упаковок бінарних сумішей дисків різного розміру. Упаковки були побудовані з використанням моделі випадкової послідовної адсорбції (RSA) з одночасним осадженням дисків двох розмірів. Диски меншого розміру обиралися для осадження з ймовірністю $p$, а диски більшого розміру з ймовірністю $1-p$. Співвідношення діаметрів дисків змінювалося в діапазоні $D=1$--$10$, а ймовірність $p=0$--$1$. Вважалося, що зв'язність дисків в упаковці забезпечується шляхом перекриття м'яких оболонок навколо дисків (структура тверде ядро--м'яка оболонка). Проаналізовано поведінку коефіцієнта покриття поверхні дисками в стані джаммінгу, зв’язності упаковок по малих, великих та обох типах дисків, а також поведінку електропровідності плівок для суміші малих провідних і великих непровідних дисків. Перколяційна поведінка електропровідності упаковки пояснена утворенням провідних ``містків'' з малих дисків у порах між великими дисками.

\keywords упаковка дисків, заклинення, адсорбція, конкуренція, перколяція

\end{abstract}

\end{document}